\documentclass[12pt]{iopart}

\usepackage{iopams}  
\usepackage{graphicx}
\newcommand*{\ket}[1]{\mathopen{|}#1\mathclose{\rangle}}

\begin{document}

\title{Connecting entanglement renormalization and gauge$/$gravity dualities}

\author{Javier Molina-Vilaplana}

\address{Department of Systems Engineering and Automation. Technical University of Cartagena. 
C/Dr Fleming S$/$N. 30202. Cartagena. Spain}
\ead{javi.molina@upct.es}
\begin{abstract}
In this work we provide additional support for the proposed connection between the gauge/gravity dualities in string theory and the successful Multi-Scale-Entanglement-Renormalization-anstaz (MERA) method developed for the efficient simulation of quantum many body systems at criticality. This support comes by showing an explicit formal equivalence between the real space renormalization group (RG) flow of the two point correlation functions in different types of MERA states and the holographic RG flow of these correlation functions in asymptotically Anti de Sitter (AdS) spacetimes. These observations may be useful in order to formalize and make more precise the connection between the properties of different MERA states and their potential holographic descriptions.
\end{abstract}

\maketitle

\section{Introduction}
According to the holographic principle, in its most general form, it is possible to establish a correspondence/duality between physics defined in a space of dimension $d$ with a (generally different) physics on the boundary of that space, which has smaller dimension. The best known concrete realization of the holographic principle is the AdS/CFT correspondence \cite{adscftbib}. In its original formulation, this duality establishes a correspondence between an $\mathcal{N}=4$ supersymmetric Yang-Mills strongly coupled four-dimensional gauge field theory (which turns out to be a conformal field theory, CFT) and a type IIb superstring theory living in the five-dimensional anti-de Sitter space (AdS$_5\, \times S_5$). Under certain conditions (the number of gauge field flavors to be large) the theory of superstrings turns out to be a classical gravitational theory of weakly interacting particles. Therefore, a strongly-coupled quantum field theory on one side of the correspondence is described by a classical theory of (super)gravity on the other side. While the former member of duality is difficult to deal with, the latter one is rather easily treatable.

Since the initial formulations of the correspondence, the AdS/CFT duality has been much generalized and has provided numerous valuable simple models to study non-perturbative  effects such as confinement and quantum phase transitions. However, despite a huge amount of work, the microscopic roots of the duality are not well understood. For instance, there is not a first principles derivation of the correspondence and the question about whether a bulk theory can be shown to be dual to a boundary theory without invoking the string-theoretic roots of AdS/CFT remains open \cite{douglas11}. In this sense, it is widely accepted that AdS/CFT is at heart a geometric reformulation of the renormalization group (RG), in which the renormalization scale becomes an extra radial dimension \cite{verlindes, sakai03}.

Recently, the outstanding progress in the setting up of real space quantum renormalization group methods based on ideas coming from quantum information, has been used to define a generalized notion of holography inspired by the AdS/CFT duality \cite{swingle09}. The author in \cite{swingle09}, has made the groundbreaking observation that the tensor networks of the \emph{Multi Scale Entanglement Renormalization Ansatz} (MERA) real space renormalization group procedure \cite{vidalmera, merait}, happens to be a realization of the AdS/CFT correspondence. The MERA procedure is implemented through a layered network of isometric tensors, where each layer defines a RG transformation: prior to the renormalization of contiguous blocks of typically two or three lattice sites into a single site, short range entanglement between the blocks is removed. The MERA coarse-graining transformation induces an RG map that can be applied arbitrarily many times keeping constant the computational cost for obtaining the consecutive effective theories yielded by the method \cite{Evenbly10}. This renders this procedure well suited to the study of quantum critical systems, where the MERA tensor network simplifies, and acquires a characteristic scale-invariant structure \cite{meracft}. The tensor networks in MERA organize the quantum information contained in a state in terms of different scales and according to \cite{swingle09}, the dual higher dimensional geometry emerges when one realizes that the tensors in a one dimensional scale invariant MERA related with a quantum critical point, are connected so as to represent a discrete version of the three dimensional anti-de Sitter space (AdS$_3$). These ideas about the holographic structure of MERA have been subsequently developed/extended in \cite{Evenbly11}. Namely, the decay of correlations and the scaling of the entanglement entropy given by MERA, may be explained in terms of the geodesics and the minimal surface regions appearing in the holographic proposal of \cite{ryu}. Using this, the holographic structure of MERA has been exploited to analyze the correlations between two disjoint regions of a critical system described by a ($d + 1$) dimensional CFT lying at the boundary of an asymptotically AdS$_{d+2}$ spacetime \cite{Molina11}.
 
In this work we provide further support for the connection between the powerful toolbox of gauge/gravity dualities and the successful entanglement renormalization numerical MERA method posed in \cite{swingle09, Evenbly11} by explicitly observing the correspondence between the RG flow of the two point correlations in MERA and the RG flow of these functions in the context of the holographic renormalization group \cite{verlindes, sakai03}. The holographic RG is based on the idea that the radial coordinate of a space-time with asymptotically AdS geometry can be identified with the RG flow parameter of the boundary field theory. This interpretation is rather non-trivial because defining a real space quantum renormalization group is not straightforward. In light of this, it is thus interesting to explicitly use entanglement renormalization for making more precise the role of the real space renormalization group in holography.

The paper is organized as follows: in section 2 we review the equations accounting for the RG flow of the two point functions in the scale invariant MERA states and then proceed to explicitly show in section 3, that these equations are equivalent to those arising when considering the holographic RG of two point functions in asymptotically AdS$_{3}$ spacetimes.  In section 4 we extend our analysis and argue that for a particular choice of finitely correlated MERA states, there is a reasonable holographic interpretation in terms of an AdS$_3$ black hole. A discussion of arguments showed in this paper is presented in section 5. Finally, we summarize our results.    

\section{Correlators in Scale Invariant MERA}
Each layer in a MERA tensor network implements a RG transformation which generates an individual step in a sequence of coarse grained lattices, $\mathcal{L}_0  ~\rightarrow  ~\mathcal{L}_1 ~\rightarrow ~\cdots ~\rightarrow ~\mathcal{L}_{w-1} ~\rightarrow ~\mathcal{L}_{w} ~\rightarrow ~\cdots$. If $\mathcal{O}$ denotes a local observable defined on one site of our original lattice $\mathcal{L}_0$, then in an scale invariant MERA, the \emph{one site scaling superoperator} $\mathcal{S}^{(1)}$ is defined by the operation $\mathcal{O}_w = \mathcal{S}^{(1)}(\mathcal{O}_{w-1})$ and reads \cite{meracft},
 
\begin{equation}
\mathcal{S}^{(1)}(\circ) = \sum_{\alpha} \lambda^{(1)}_{\alpha} \phi^{(1)}_{\alpha} \rm{Tr}(\phi^{(1)}_{\alpha}\circ), 
\label{spectral1}
\end{equation}

where the scaling dimensions of the scaling operators $\phi^{(1)}_{\alpha}$ of the theory are $\Delta^{(1)}_{\alpha}\equiv -\log \lambda^{(1)}_{\alpha}$ and $\rm{Tr}(\phi_{\alpha} \phi_{\beta}) = \delta_{\alpha\beta}$. The correlator $C_{\alpha\beta}^{w}$ of two scaling operators $\phi_{\alpha}^{(1)}$ and $\phi_{\beta}^{(1)}$ placed on contiguous sites on the $w$-th level of the MERA tensor network reads,
\begin{equation}
C_{\alpha\beta}^{w} \equiv \left\langle \phi_{\alpha}^{(1)}(1) \phi_{\beta}^{(1)}(0)\right\rangle. 
\label{C2}
\end{equation}

It is worth to point out that $\phi_{\alpha}^{(1)}$ and $\phi_{\beta}^{(1)}$ are initially placed in the sites $x$ and $y$ of the original lattice $\mathcal{L}_0$. Thus, after $w \sim \log \vert x -y \vert$ iterations of the RG transformation, $\phi_{\alpha}^{(1)}$ and $\phi_{\beta}^{(1)}$ become first neighbors (see Eq.(\ref{C2})) \cite{meracft} though, strictly speaking, $w=\log_2 \vert x-y\vert$ or $w=\log_3 \vert x-y \vert$ depending on the MERA implentation, binary or ternary repectively. Each RG iteration contributes a factor $\lambda^{(1)}_{\alpha}\lambda^{(1)}_{\beta}$, so

\begin{eqnarray}
C_{\alpha\beta}^{w} & = \left\langle \mathcal{S}^{w}(\phi_{\alpha}(x)) \mathcal{S}^{w}(\phi_{\beta}(y))\right\rangle \label{scaling3}\\
& =(\lambda_{\alpha} \lambda_{\beta})^{w}\left\langle \phi_{\alpha}(x) \phi_{\beta}(y)\right\rangle \\
& = (\lambda_{\alpha} \lambda_{\beta})^{w} C_{\alpha\beta}^{0} , 
\end{eqnarray}

where the superscript $-1-$ referring to one site operators has been dropped out and $C_{\alpha\beta}^{0}\equiv \left\langle \phi_{\alpha}(x) \phi_{\beta}(y)\right\rangle$. Defining the change of variable $w=\log z$, taking into account that $\Delta_{\alpha, \beta}\equiv -\log \lambda_{\alpha, \beta}$ and the identity $a^{\log b}= b^{\log a}$, the flow of $C_{\alpha\beta}$ along consecutive MERA RG steps (now labelled by the variable $z=e^{w}$) can be written as,
\begin{equation}
 C_{\alpha \beta}^{z}=z^{-\eta} C_{\alpha \beta}^{1},
 \label{merascaling}
\end{equation}

where $\eta=(\Delta_{\alpha} + \Delta_{\beta})\delta_{\alpha\beta}=2\Delta_{\alpha}$. Defining $\partial_z \equiv \partial / \partial z$, one clearly sees that the equation (\ref{merascaling}) is the solution of the differential equation,
\begin{equation}
\left( z\, \partial_{z}   + \eta \right) \, C^{z}_{\alpha\beta} = 0,
\label{csmera}
\end{equation}

which provides the MERA RG flow of the two point correlation functions $C_{\alpha \beta}^{z}$. 

\section{Holographic RG in the AdS geometry}
In order to provide some explicit formal link between MERA and holography, we analyze the RG flow of the two point functions in the AdS$_{3}$ spacetime. As we are interested in time-independent ground states, the time dimension is irrelevant in our discussions. Thus, we consider a generic metric where a dimension $x$ is fibered in a nontrivial way over the holographic coordinate $w$, such as,
\begin{equation}
ds^2 = a^2(w)dx^2 + dw^2.
\label{metric1}
\end{equation}

The function $a(w)$ is called the warp factor (also known as Weyl factor) which assumes $a(w)=e^{-w}$ when considering AdS space. Using the change of variable $w=\log z$, one obtains,
\begin{equation}
ds^2 = \frac{1}{z^2}(dx^2 + dz^2),
\label{metric2}
\end{equation}

and the warp factor turns into $a(z)=1/z$. In \cite{verlindes, cGomez} it is shown that the holographic renormalization group equations for correlators $C$ (Callan-Symanzik equations) related with the geometry (\ref{metric1} - \ref{metric2}) can be written as,
\begin{eqnarray}
\left(a\, \partial_{a}  - \beta\, \partial_{\Phi}\right)C = 0, 
\label{holocs}
\end{eqnarray}

where $a$ is the warp factor $a(z)=z^{-1}$, $\beta \partial_{\Phi}  \equiv a  \partial_{a} \Phi$,
 and $\Phi(x,z)$ is the solution to the classical Einstein equations of motion for massive free fields moving in the  metric (\ref{metric2}) with the boundary condition $\Phi(x)\equiv \Phi(x,z=0)$. Equation (\ref{holocs}) in terms of the holographic dimension $z$ reads as,
\begin{equation}
\left(z\, \partial_{z}  - z\, \partial_{z}\Phi\, \partial_{\Phi} \right)C = 0. 
\label{holocs2}
\end{equation}
 In the holographic duality dictionary \cite{adscftbib}, the scaling dimensions $\Delta_{\alpha}$ of the boundary operators $\phi_{\alpha}$ are related with the masses $m$ of their dual fields $\Phi$ in the bulk geometry. This relation can be derived by solving the field equation for $\Phi$, 
 
\begin{equation}
  \left(-\Box + m^2 \right) \Phi = 0,
  \label{EoM}
\end{equation}

  where $\Box$ is the \emph{Laplacian} operator defined by,
 
\begin{equation} 
  \Box \Phi \equiv \frac{1}{\sqrt{g}} \partial_{A} \left(\sqrt{g} g^{AB}\partial_{B}\right)\, \Phi \,
  \label{laplacion}
  \end{equation}
  
   with $A, B = \left\lbrace x, z \right\rbrace$, $g^{AB}$ the metric defined in (\ref{metric1}-\ref{metric2}), $g \equiv \det{\left(g^{AB}\right)}$ and imposing that the radial dependence of $\Phi$ is $\Phi(z)\sim z^{-\Delta_{\alpha}}$ \cite{adscftbib}. If we use this \emph{ansatz} and noticing that $C\equiv\left\langle \Phi \Phi \right\rangle \sim z^{-2\Delta_{\alpha}}\sim \Phi^{2}$, then equation (\ref{holocs2}) can be written as,
 \begin{equation}
 \left(z\, \partial_{z}  +  2\Delta_{\alpha}\right)C = 0,
 \label{holocs3}
 \end{equation}
 
that is exactly the differential equation (\ref{csmera}) for the MERA flow of correlation functions. The equivalence between (\ref{csmera}) and (\ref{holocs3}) substantiates, in our opinion, the proposal that the tensor network structure of an scale invariant MERA may be suitably described by the \emph{renormalization group interpretation} of a higher dimensional gravity dual corresponding to the $\rm{AdS}_{3}$ space-time \cite{swingle09, Evenbly11}, further aspects of which we will explore below. 
 
 One could also have derived this result by means of the following: from (\ref{merascaling}) one clearly sees that, as one moves deeper along the holographic direction $z \to z^{'}$,  $C^{z}_{\alpha \beta}$ scales as,
 \begin{equation}
 C_{\alpha \beta}^{z^{'}}=(z^{'}/z)^{-\eta}\,  C_{\alpha \beta}^{z},
 \label{corrscale}
 \end{equation}
where $z^{'}=z e^{u}$ corresponds to a displacement in the direction of the RG coordinate $w$ ($w^{'}=w+u$). Indeed, correlation functions $C_{\alpha \beta}$ on different depths of the holographic coordinate $z$ are related by the simple rescaling  $(z^{'}/z)^{-\eta}$. This is exactly the behaviour expected for the long distance behaviour of the correlation functions in a Wilsonian effective treatment. A Wilsonian renormalization procedure amounts to integrate out short distance degrees of freedom and absorb the result into varying coupling constants which are therefore said to flow. This procedure shows that long distance physics is not sensitive to short distance details \cite{wilson74}. Both MERA and a Wilsonian RG procedure carry out a covariant reescaling of the physical coordinates ($x$ in our considerations). Thus, if one does not merely interpret $z$ as a variable which measures the depth in the MERA tensor network, but otherwise, adopts it as a new coordinate in a geometrical dual description of the MERA RG flow, then the Wilsonian interpretation of (\ref{corrscale}) leads to postulate a metric for this dual geometric model with a factor accounting for the covariant rescaling of the coordinate $x$. This is precisely the effect of the warp factor $a(z)=1/z$ in (\ref{metric2}). Put otherwise, the metric that describes the characteristic scaling of the two-point correlation functions within the MERA framework, corresponds to the metric of the $\rm{AdS}_3$ space \cite{kraussbalb} when (\ref{corrscale}) is looked at from a Wilsonian viewpoint. Namely, it is easy to check that (\ref{metric2}) is invariant under the Wilsonian covariant rescaling $z \to z e^{u}$ and $x \to x e^{u}$. 
 
\section{Truncated holographic geometry for finite range MERA states}
In the AdS/CFT \cite{adscftbib} correspondence, a given geometry corresponds to some state in the dual field theory. For instance, massive two dimensional quantum field theories are represented in the dual gravity side by an infrared (IR) deformation of the AdS$_{3}$ space that caps off or truncates the IR region $z > z_{IR}$.  Thus, the important quantity in the massive theory is the correlation length $\xi$, which is identified with $\xi \sim z_{IR}$ in AdS/CFT \cite{ryu}.

In \cite{swingle09, Evenbly11}, the emergence of the gravity dual picture to a MERA tensor network state is intimately related to the structure of the quantum entanglement between the microscopic degrees of freedom of the quantum system under consideration \cite{vanram09}. Indeed, this amount of entanglement depends on the scale at which one is probing the system. For instance, a gapped state presents no correlations beyond a distance of order of the inverse mass. Similarly, a deconfined finite temperature system at temperature $T$ will also present no correlations for distances larger than a characteristic length scale $\xi = 1/T$.

Let us then consider the ground state $\ket{\Psi_{\rm{GS}}} \equiv \ket{\Psi(\mathcal{L}_{0})}$ of a system, in which the two point correlation functions decay exponentially due to the existence of a characteristic length scale $\xi$. In this case, the correlation length $\xi$ shrinks to one (in units of separation between lattice sites) after $w_{*} \sim \log \xi$ coarse-graining steps. In other words, one needs $w_{*} \sim \log \xi$ renormalization steps to obtain a coarse grained version $\ket{\Psi(\mathcal{L}_{w_{*}})}$ of the ground state under consideration in which all two point correlators decay exponentially. Following \cite{Evenbly11}, it is thus reasonable to assume that this later state may be well approximated by the product state

\begin{equation}
\ket{\Psi(\mathcal{L}_{w_{*}})} \equiv \ket{0}\otimes \ket{0} \otimes \cdots \otimes \ket{0},
\label{prodstate}
\end{equation}

(\textit{i.e} a state with no correlations between different lattice sites) and that the state $\ket{\Psi_{\rm{GS}}}$ may be represented by a \emph{finite range} MERA, composed by a finite number $w_*$ of MERA layers and which geometrical description results truncated in the holographic RG direction by the presence of a \emph{truncating boundary} in the value of the scale parameter $z = e^{w}$ given by $z_{*} \equiv \xi \sim z_{IR}$. 

In order to extend the explicit link between MERA and gauge/gravity dualities of the previous section, in the following, we will consider a particular choice of finite range MERA, in which its first $w_{*}$ layers of tensors correspond to those in the scale invariant MERA that describes the neighbouring critical point. Here we try to show, by means of an holographic argument, that this finite range MERA may suitably represent the ground state of an infinite scale invariant quantum critical system heated up to the finite temperature $T=1/\, \xi$.

With this choice of finite range MERA, let us now consider the correlator between two scaling operators $\phi_{\alpha}$ and $\phi_{\beta}$ located at sites $s_1$ and $s_2$ in   $\mathcal{L}_{0}$ and separated by a distance $r = \vert s_1 - s_2\vert$ such that $r\ll\xi$. In MERA, the \emph{causal cone} of a set of sites located at $\mathcal{L}_0$ is defined by all the tensors in $\left\lbrace \mathcal{L}_1, \cdots \mathcal{L}_w, \cdots\right\rbrace$ that affect the expectation values of operators with support on those sites \cite{vidalmera}. When the causal cones of two sites merge at some level $w$ of the scale invariant tensor network, the operators defined on the boundary are correlated with the algebraic decaying functional dependence of (\ref{merascaling}). When $r\ll\xi$, the causal cones of the two scaling boundary operators $\phi$ overlap after $w \sim \log r$ MERA steps. As a result, as $w \ll w_{*}$, the stereotypical $\Lambda$-shaped causal cone $\mathcal{C}^{[s_1,s_2]}$ (Fig1, left) is not distorted by the presence of the boundary at $z_{*}$. Namely, the MERA RG flow only runs through length scales $z \ll z_*$ while not being affected by the truncating boundary at $z=z_*$ and giving for the two point function of the two scaling operators,

\begin{equation}
C_{\alpha\beta}^{z} \equiv \left\langle \phi_{\alpha}(1) \phi_{\beta}(0)\right\rangle = z^{-\eta} C_{\alpha \beta}^{1},
 \label{asympads}
 \end{equation}
 
with $z=r$. As posed in \cite{swingle09, Evenbly11, Molina11}, the boundary of the causal cone in an optimized scale invariant MERA is a curve of minimal length (geodesic) connecting the boundary points $s_1$ and $s_2$ in the dual AdS description of the tensor network. In the case of $z \ll z_*$, the geodesic connecting the two sites within the truncated holographic geometry only probes the $\rm{AdS}_{3}$ region of the finite range MERA geometry that we are considering in this section.
  
 At variance, when $r\gg\xi$, the geodesic curve connecting $s_1$ and $s_2$ within the tensor network (the length of the curve which is the boundary of $\mathcal{C}^{[s_1,s_2]}$) results distorted (namely, it grows proportional to $r$, Fig1, right) as a consequence of the truncation of the $\Lambda$-shaped causal cone $\mathcal{C}^{[s_1,s_2]}$ at $z_{*}$ \cite{Evenbly11}. As posed by authors in \cite{Evenbly11}, while it is tempting, it is not straightforward to infer a justified scaling for $C^{z}_{\alpha \beta}$ when $z \gg \xi$ by directly looking at the geometry of the truncated causal cone. Nevertheless, some intuition may point out to expect an exponential decay of the correlators. Namely, a finite range MERA in $d=1$ dimensions composed of $w_* \approx \log \xi$ layers of tensors and bond dimension $\chi_{\rm{MERA}}$, can be re-expressed as a matrix product state (MPS) with bond dimension $\chi_{\rm{MPS}}$ given by
   
\begin{equation}
	\chi_{\rm{MPS}} \approx \left(\chi_{\rm{MERA}}\right)^{w_*}.
	\label{MPSvMERA}
\end{equation}

Our intuition is confirmed as one realizes that in an infinite and translational invariant MPS with finite bond dimension $\chi_{\rm{MPS}}$, the correlation between two sites decays exponentially with the distance $r$ between them \cite{Hastings07} and thus, a sensible \emph{ansatz} for the scaling of 
$C^{z}_{\alpha \beta}$ when $z \gg z_{*}$ is given by,

\begin{equation}
C_{\alpha\beta}^{z} \equiv \left\langle \phi_{\alpha}(1) \phi_{\beta}(0)\right\rangle \approx  e^{ -\eta\, \left(z/z_{*}\right)}.
\label{decayexp}
\end{equation} 

The two limits $z \ll z_*$ (Eq (\ref{asympads})) and $z \gg z_*$ (Eq. (\ref{decayexp})) for the scaling of $C^{z}_{\alpha \beta}$ in the finite range MERA that we are considering in this section, may be expressed in terms of the mixed form,
\begin{equation}
C_{\alpha\beta}^{z} \equiv \left\langle \phi_{\alpha}(1) \phi_{\beta}(0)\right\rangle \approx z^{-\eta}\, e^{-\eta\, (z/z_{*})}\, C_{\alpha\beta}^{1},
\label{finitcorr}
\end{equation} 

in which the algebraic decay of correlations dominates for distances smaller than the correlation length, $|s_1-s_2| \ll \xi$, and the exponential decay of Eq.(\ref{decayexp}) is the leading contribution at distances such that $|s_1-s_2| \gg \xi$. This is precisely the scaling of the two point functions that one should expect for a non-critical system that is close to a quantum critical point \cite{DiFrancesco97}, and suggests that our particular choice of the finite range MERA, may suitably represent a massive state that is close to a quantum critical point.
 
 \begin{figure} 
\begin{center}
\includegraphics[width=5.25 in, height=2.75 in]{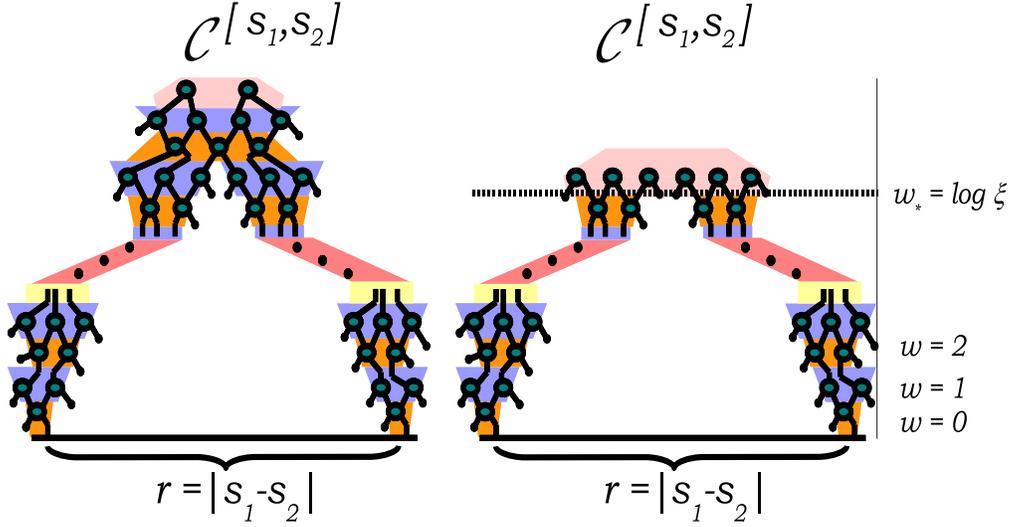} 
\end{center}
\label{Fig1}
\caption{Causal cones ${\cal C}^{[s_1s_2]}$ for two one site scaling operators $\phi$
in an scale invariant MERA (left) and in a the finite range MERA in which its first $w_{*}$ 
layers of tensors correspond to those in the scale invariant MERA that describes
the neighbouring critical point (right).}
\end{figure}

 Our aim now, consists in giving an appropiate holographic dual model for the state that we are considering. To this end, let us first express the scaling of $C_{\alpha\beta}^{z}$ for $z\gg z_{*}$ in Eq. (\ref{decayexp}) as,
 \begin{equation}
C_{\alpha\beta}^{z} = C_{\alpha\beta}^{z_{*}}\,  e^{ -\eta\, \left(z/z_{*}\right)},
\label{decayexp2}
\end{equation} 
where $C_{\alpha\beta}^{z_{*}}$ is a constant that for convenience, we assume that depends on $z_*$. Here, it is pertinent mentioning that Eq. (\ref{decayexp2}) is the solution of,
 
 \begin{equation}
 \left(z\, \partial_{z}  + \eta\, (z/z_{*})\right)\, C^{z}_{\alpha \beta} = 0,
 \label{CSfincorr}
 \end{equation}
  
 which provides, assuming Eq.(\ref{decayexp2}), the MERA RG flow for $C^{z}_{\alpha \beta}$ when $z \gg z_*$. Equation (\ref{CSfincorr}), may be derived using the \emph{ansatz} $\Phi(z)\sim \exp [-\Delta_{\alpha}\, (z/z_{*})]$ in (\ref{holocs2}). Namely, this amounts to consider that,
 \begin{equation}
C\equiv\left\langle \Phi \Phi \right\rangle \sim \exp [-2\Delta_{\alpha}\, (z/z_{*})]. 
\label{scalansatz}
\end{equation}
  
 In terms of the AdS/CFT dictionary \cite{adscftbib}, Eq. (\ref{scalansatz}) corresponds to the propagator of a massive scalar field in an AdS$_3$ geometry which includes a black hole with the event horizon located at a distance $z_{*} = \xi$ from the boundary of the space (\textit{i.e} at $z\ll z_{*}$). 
 
 The AdS black hole is an asymptotically AdS spacetime with metric,
\begin{equation}
ds^2 = \frac{1}{z^2}\left( dx^2 + \frac{dz^2}{f(z)}\right),
\label{bhmetric}
\end{equation}
where $f(z)$ is the emblackening factor,

\begin{equation}
 f(z)=1-\left(\frac{z}{z_*}\right)^{2}.
 \label{emblack}
 \end{equation}
 
  According to the AdS/CFT correspondence, an AdS black hole is dual to a thermal state at finite temperature $T = 1/z_*$ on the conformal boundary. One may regard this black hole as a small perturbation of the boundary CFT. The AdS$_3$ black hole (also known as BTZ black hole, \cite{btz}) without any electric charge is locally isometric to anti de Sitter space. More precisely, it corresponds to an orbifold of the universal covering space of AdS$_3$. 

 In the following, we will show that the scaling of the correlation functions given in equations (\ref{asympads}) and (\ref{decayexp2}) may be derived by considering the propagator of massive fields in the dual geometry with the metric given by (\ref{bhmetric}). First, we note that the AdS/CFT correspondence \cite{adscftbib} states that the propagator of a massive scalar field in AdS between two points $\bf{b_1}$ and $\bf{b_2}$ located at an arbitarily small distance $\epsilon$ from the points $s_1$ and $s_2$ in the boundary (\textit{i.e} at $\mathcal{L}_{0}$ in MERA) is given by,

 \begin{equation}
\left\langle \phi_{\alpha}(\bf{b_1}) \phi_{\beta}(\bf{b_2})\right\rangle_{\bf{Holo}} \sim \exp \left(-\eta\, \mathcal{D}_{g}\left[ \bf{b_1},\bf{b_2} \right]\right).
 \label{holopres} 
 \end{equation}
 
 Here, $\mathcal{D}_{g}\left[ \bf{b_1},\bf{b_2} \right]$ is the geodesic distance in the metric (\ref{bhmetric}) between the points $\bf{b_1}$ and $\bf{b_2}$. We stress that the holographic prescription (\ref{holopres}) is consistent since asymptotically AdS metrics such as (\ref{metric2}) and (\ref{bhmetric}) diverge as $z \to 0$ \textit{i.e}, all the computations remains finite if one assumes that that the propagation of the dual fields $\Phi$ occurs for $z \geq \epsilon$ with $\epsilon$ being arbitrarily small. The geodesic distance $\mathcal{D}_{g}\left[ \bf{b_1},\bf{b_2} \right]$ is therefore given by \cite{adscftbib, btz},
 
 \begin{equation}
 \mathcal{D}_{g}\left[ \bf{b_1},\bf{b_2} \right]= \log \left[z_{*}\, \sinh\left( \frac{z}{z_{*}}\right)   \right], 
 \label{dist}
 \end{equation}
 
 with $z=\vert \bf{b_1} - \bf{b_2}\vert$ and $z_{*}=\xi=1/T$. Substituting (\ref{dist}) into (\ref{holopres}) yields,
 
 \begin{equation}
 \left\langle \phi_{\alpha}(\bf{b_1}) \phi_{\beta}(\bf{b_2})\right\rangle_{\bf{Holo}} \sim \left[z_{*}\, \sinh\left( \frac{z}{z_{*}}\right)   \right]^{-\eta}. 
 \label{holocorr}
 \end{equation} 

Let us first analize the behaviour of (\ref{holocorr}) when $z=\vert \bf{b_1} - \bf{b_2}\vert$ $\ll z_*$. For this regime, the $\rm{AdS}_3$ metric (\ref{metric2}) is recovered from Eq.(\ref{bhmetric}) as the emblackening factor $f(z) \to 1$ while the length of geodesics connecting $\bf{b_1}$ and $\bf{b_2}$ is given by,

\begin{equation}
\mathcal{D}_{g}\left[ \bf{b_1},\bf{b_2} \right]\approx \log (z).
\label{distmod1}
\end{equation}

\begin{figure}
\begin{center}
\includegraphics[width=5.5 in, height=2.25 in]{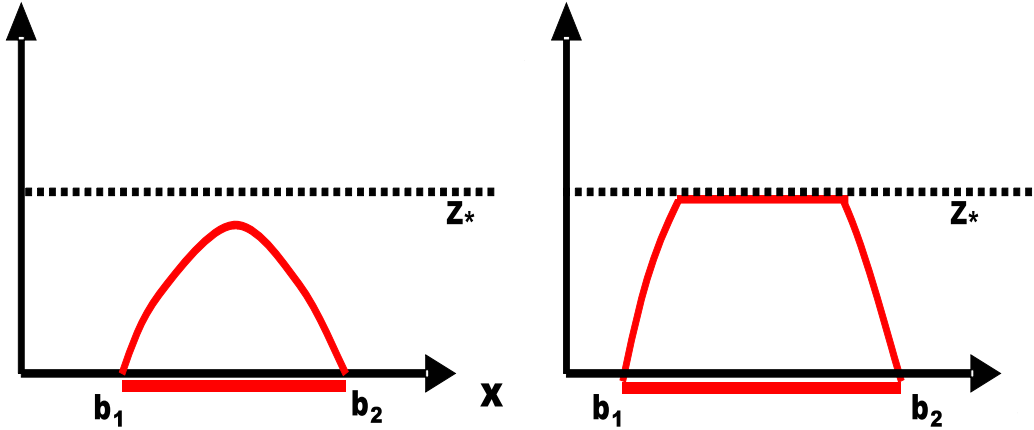} 
\end{center}
\label{Fig2}
\caption{Geodesic curves (in red) that appear when computing the holographic correlator Eq.(\ref{holopres}) of the scaling operator $\phi$ located at points $\bf{b_1}$ and $\bf{b_2}$ in the AdS black hole geometry. Left: When $z = \vert \bf{b_1} - \bf{b_2} \vert$ $\ll z_*$, the geodesic curve does not probe the near horizon region ($z \sim z_*$) (dotted line). Right: When $z \gg z_*$, the geodesic wraps around the horizon so the length of the curve now scales $\propto \vert \bf{b_1} - \bf{b_2} \vert$ giving rise to an exponential decay of the correlations.}
\end{figure}
 
 By substituting (\ref{distmod1}) into (\ref{holocorr}) one obtains that,
  
 \begin{equation}
 \left\langle \phi_{\alpha}( \bf{b_1}) \phi_{\beta}( \bf{b_2})\right\rangle_{\bf{Holo}} \approx z^{-\eta},
 \label{holores1}
 \end{equation}
 
 that is equivalent (up to a constant) to (\ref{asympads}). This corresponds to the situation where the separation between the scaling operators $\phi_{\alpha}$ and $\phi_{\beta}$ in the boundary of the AdS$_3$ spacetime is not large enough for the geodesic connecting $\bf{b_1}$ and  $\bf{b_2}$ to probe the region near the horizon of the black hole (Fig 2, left). 
 
 On the other hand, when $z=\vert \bf{b_1} - \bf{b_2}\vert$ $\gg z_{*}$, the length of the geodesic connecting the two sites within the tensor network grows proportional to $\vert \bf{b_1} - \bf{b_2}\vert$ and the behaviour of $\left\langle \phi_{\alpha}( \bf{b_1}) \phi_{\beta}( \bf{b_2})\right\rangle_{\bf{Holo}}$ is given by,
  
 \begin{equation}
 \left\langle \phi_{\alpha}( \bf{b_1}) \phi_{\beta}( \bf{b_2})\right\rangle_{\bf{Holo}} \approx z_{*}^{-\eta}\, e^{-\eta\, (z/z_{*})},
 \label{holores2}
 \end{equation}
 
which corresponds to the case in which the geodesic curve connecting $\bf{b_1}$ and  $\bf{b_2}$ probes the region near the horizon of the black hole ($z \sim z_*$, Fig 2, right). This yields an exponential decay of the correlations which might be identified with the decay in Eq.(\ref{decayexp2}) if one assumes that $C_{\alpha \beta}^{z_*}\equiv (z_*)^{-\eta}$.

 Comparing the equations (\ref{asympads}) and (\ref{decayexp2}) with (\ref{holores1}) and (\ref{holores2}) respectively, allows us to write that,
  
   \begin{equation}
 \left\langle \phi_{\alpha}( \bf{b_1}) \phi_{\beta}( \bf{b_2})\right\rangle_{\bf{Holo}} \sim \frac{C^{z}_{\alpha \beta}}{ C^{1}_{\alpha \beta}}\, .
 \label{holMERA}
 \end{equation}
 
  According to the AdS/CFT correspondence \cite{adscftbib} and the proposals in \cite{swingle09, Evenbly11}, the equivalence between the correlators of two scaling operators $\phi$ in the type of finite range MERA that we have been considering here and the holographic propagators of massive fields $\Phi$ in the AdS$_3$ black hole geometry, lead us to conclude that, these type of MERA states represent the ground state of a CFT at the finite temperature $T=1/\, \xi$. 
 
 \section{Discussion}
The arguments presented in the previous sections are based on an explicit analysis of the behaviour of two point functions in two types of MERA tensor states. It is interesting to discuss some features of the conclusions raised there from different points of view. First, let us to briefly analyze if our interpretation of the finite range MERA composed by $w_* \approx \log \xi$ layers of tensors which are associated to an scale invariant MERA, in terms of the vaccum state of a CFT at the finite temperature $T=1/\, \xi$, is consistent with the known results for the scaling of the entanglement entropy in these states \cite{Calabrese04}. Following \cite{swingle09}, the higher dimensional geometry defined by MERA may be usefully visualized by locating cells around all the sites of the tensor network: these cells are unit cells filling up the emerging geometry and the size of each cell is defined to be proportional to the entanglement entropy of the site in the cell. The scale invariance of the first $w_*$ layers in our finite range MERA forces each layer to be identical and all the sites at each layer to be equivalent and have the same size. 

In MERA, as one proceeds to compute the entanglement entropy of a block of $\ell$ sites in $\mathcal{L}_0$, it is necessary to trace out any site in the bulk geometry defined by the tensor network which does not lie in the causal cone of the block. The boundary of this causal cone is a curve in the MERA induced AdS higher dimensional geometry which length is, by definition, the sum of the entropies of all the traced out sites. Namely, in an optimized finite range MERA similar to the one considered here, the length of this geodesic curve is given by (\ref{dist}) where now, $z \equiv \ell$ \cite{swingle09}. 

If one considers a block of size such that $\ell \ll z_*$, in this case, the geodesic in the truncated holographic geometry of our finite range MERA state, only runs through length scales $z \sim \ell$ smaller than $z_*$, \textit{i.e} the geodesic only runs through the scale-invariant AdS$_3$-like region of the finite range MERA and its length remains logarithmic in $\ell$ as in a critical system. This yields the well known scaling for the entanglement entropy of the block,

\begin{equation}
S \sim \log \ell.
\label{ent1}
\end{equation}

 On the other hand, when $\ell \gg z_*$, the length of the geodesic curve grows proportional to $\ell$ so,
 
\begin{equation}
S \sim \frac{\ell}{z_*} + \log z_*\, . 
\label{ent2}
\end{equation} 

  In other words, as the spatial size of the filling cells of the sites in the boundary of the causal cone of the block are all non zero, the sum of the entropies associated to these sites become proportional to the size of the block. As a consequence, the entanglement entropy of a block of size $\ell \gg 1/\, T$ in our finite range MERA, has the usual boundary contribution plus an extensive piece, consistent
with the presence of a black hole horizon \cite{ryu}.

Note that this interpretation does not hold for all finite range MERA states. Namely, the tensors in different layers of finite range MERA states are generally different, which is a manifestation of the fact that the amount of entanglement in the ground state of these states varies depending on the length scale investigated. For a general finite range MERA state, the size of the filling cells at the initial layers of the network is non-zero because there is short range entanglement in the system. However, entanglement and therefore, the size of the filling cells, decreases as one moves deeper into the MERA geometry (\textit{i.e}, as one probes bigger length scales in the system) until it is absent at a layer of the MERA network called \emph{factorization scale}, which is equivalent to the characteristic length scale of the system \cite{swingle09}. The size of the filling cells at this layer shrinks to zero and thus, the entanglement entropy of a block receives contributions only from a finite range of scales while there is no trace of an extensive term in the entropy of the block.

Finally, it is worth to regard that the two different regimes for the RG flow of the two point correlation functions in the finite range MERA that has been considered here (Eq. (\ref{holocs3}) for $z \ll z_*$ and Eq.(\ref{CSfincorr}) for $z \gg z_*$), reflect the non-trivial global structure of the MERA renormalization group flow in these states. Namely, an holographic interpretation has been given in terms of an AdS black hole. In the previous section we have mentioned that the presence of a black hole in the AdS$_3$ space corresponds to an orbifolding of this space. To be more explicit, let us note that 3-d AdS gravity can be described by a Chern-Simons theory with a boundary (Wess-Zumino-Witten WZW model) and is exactly solvable \cite{town86, witten88}. This fact, allows to translate the AdS$_3$ orbifolding into the gauging of the $SL(2,\bf{R})$ isometries of the AdS$_3$: in the region where $z \ll z_*$ (asymptotic AdS region) the dual CFT is given by an $\frac{SL(2,\bf{R})}{SO(1,1)}$ WZW model and in the near horizon region ($z \sim z_*$) one gets an $\frac{SL(2,\bf{R})}{U(1)}$ WZW model. Due to the difference in the central charges of both models, no marginal deformation between them exists. So, a nontrivial renormalization group flow is expected while the complete bulk physics is fixed by these boundary CFTs. However, it is worth to stress that the BTZ black hole geometry is locally AdS$_3$ at any point so, these two different central charges explicitly point out to the  non-trivial global structure of the underlying renormalization group flow \cite{adsorbs}.

\section{Conclusions} 
 In this paper we have provided an explicit formal equivalence between the RG flow of the two point correlation functions in two types of MERA states, and the holographic RG flow of these correlation functions in AdS spacetimes to further elucidate and make more precise the connection between holography and entanglement renormalization proposed in \cite{swingle09, Evenbly11}. 

This equivalence substantiates, in our opinion, the proposal that the tensor network structure of the MERA states under consideration may suitably be described by the renormalization group interpretation of a higher dimensional gravity dual: the AdS$_3$ if one considers the scale invariant MERA and the BTZ black hole geometry (an AdS$_3$ black hole) in the case that one deals with a finite range MERA composed of a finite number of layers made by the same tensors that would describe an scale invariant MERA related to a neighbouring critical point. In the latter case, the non trivial RG flow of the correlation functions for the finite range MERA under consideration, is naturally expalined when one considers the BTZ black hole as an orbifold of the AdS$_3$ space.

Our observations here may give some hints on how to develop a formal procedure to explicitly build the holographic dual of some generalizations of MERA accounting for exotic critical states that show violations of the boundary area law for the scaling of the entanglement entropy \cite{Evenbly11}, which motivates a deep an subtle analysis of the RG flow related with these states. 

Finally, as posed in \cite{swingle09, Evenbly11}, the emergence of the gravity dual picture to a MERA tensor network state is intimately related to the MERA coarsesing procedure (RG flow) and by the structure of the entanglement between the microscopic degrees of freedom of the quantum system at the boundary. In light of this, it is interesting to investigate which classes of coarsening transformations lead to RG flows on which there is a natural geometry.

 \ack{JMV thanks H. Wichterich for many very fruitful correspondence at the early stage of this project and the financial support by the Spanish Office for Science FIS2009-13483-C02-02 and the Fundaci\'on S\'eneca Regi\'on de Murcia 11920/PI/09.}

\end{document}